\documentstyle[12pt]{article}
\newcounter{appendixc}
\newcounter{subappendixc}[appendixc]
\newcounter{subsubappendixc}[subappendixc]

\renewcommand{\appendix}[1] {\vspace*{0.6cm}
        \refstepcounter{appendixc}
        \setcounter{figure}{0}
        \setcounter{table}{0}
        \setcounter{equation}{0}
        \renewcommand{\thefigure}{\Alph{appendixc}.\arabic{figure}}
        \renewcommand{\thetable}{\Alph{appendixc}.\arabic{table}}
        \renewcommand{\theappendixc}{\Alph{appendixc}}
        \renewcommand{\theequation}{\Alph{appendixc}.\arabic{equation}}
        \noindent{\bf Appendix \theappendixc #1}\par\vspace*{0.4cm}}

\begin{document}        
\begin{titlepage}
\begin{flushright}
hep-ph/9704257\\
AMES-HET-97-4\\
April 1997
\end{flushright}
\vspace{0.1in}
\begin{center}
{\Large $R$-parity-violating SUSY effects and signals\\
        in single top production at the Tevatron}
\vspace{.2in}

  A. Datta$^a$, Jin Min Yang$^{a,b,}${\footnote{ On leave from Physics 
Department, Henan Normal University, China}}
, Bing-Lin Young$^{a}$,  and X. Zhang$^c$

\vspace{.2in}
\it

$^a$     Department of Physics and Astronomy, Iowa State University,\\
         Ames, Iowa 50011, USA\\
$^b$     International Institute of Theoretical and Applied Physics,\\
         Iowa State University, Ames, Iowa 50011, USA\\
$^c$     Institute of High Energy Physics, Academia Sinica, \\
         Beijing 100039, China
\rm
\end{center}
\vspace{3cm}

\begin{center} ABSTRACT\end{center}

 We discuss single top quark production via $u^i \bar d^j\rightarrow t \bar b$
at the Fermilab Tevatron in the minimal supersymmetric model with $R$-parity 
violation.
We find that within the  allowed range of coupling constants,
the lepton-number violating couplings can
give rise to observable effects when the slepton mass lies  
in a specific narrow range. 
For the baryon-number violating couplings,
the contribution to the production rate can be quite large 
in the presently allowed range of the coupling constants. 
We show that the measurement of single top
production at the upgraded Tevatron  can be used to
constrain a linear combination of 
products of the $R$-parity violating couplings.
\vfill

\end{titlepage}
\eject
\baselineskip=0.30in
\begin{center} {\Large 1. Introduction }\end{center}

Single top quark production  through 
$W$-gluon fusion[1] $g+W\rightarrow t+\bar b$,
and quark-antiquark annihilation via virtual $s$-channel $W$[2]
$u^i+ \bar d^j \rightarrow t+ \bar b$, where $i,j$ are the
generation indices, are interesting to study at the Tevatron.
In contrast to the QCD process of $t\bar t$ pair production,
the single top production
involves the electroweak interaction and can, therefore, 
be used to probe the electroweak theory.
In particular, the quark-antiquark annihilation subprocess
can be used to study models of new physics in connection with the 
top quark.

It has been shown[3] that the signal for $u^i \bar d^j\rightarrow t\bar b$  is
potentially observable at the Tevatron with 2-3 $fb^{-1}$ integrated 
luminosity, although it will be overwhelmed at the LHC 
by the large background from $t\bar t$ production plus single
tops from the $W$-gluon fusion[1]. Compared to $Wg\rightarrow t\bar b$, 
$u^i \bar d^j\rightarrow t\bar b$ has the advantage that the cross section 
can be calculated reliably because the quark and antiquark
structure functions at the relevant values of $x$ are better known 
than the gluon structure functions entering in the calculation
for the $W$-gluon fusion cross section. 
With a 30 fb$^{-1}$ data sample from Run 3 at the upgraded Tevatron
with $\sqrt s=2$ TeV, it is possible to measure the ratio of
single top production ( $u^i \bar d^j\rightarrow W \rightarrow t\bar b$) 
and Drell-Yan cross section ( $u^i \bar d^j\rightarrow W \rightarrow l\nu$)
to an accuracy of $\pm 8$\% [4]. Thus new physics effects
that produce larger  than 16\% effect 
on the cross section ratio should be detectable[4].

In the Standard Model, the cross section of single
top quark production $u^i \bar d^j\rightarrow t \bar b$ has been
calculated to one loop order[5]. 
Recently, this process has been used to study the effects of 
model-independent new physics involving
the third-family quarks[6] 
and it was shown that at the upgraded Tevatron it can
be a powerful probe of new physics. 
Effects in specific new physics models on this production
process were evaluated in Refs.[7,8]. In the general two-Higgs-doublet
model and the minimal supersymmetric model (MSSM) with
$R$-parity conservation[7,9,10], 
the contribution to the production rate
can reach the observable level only for $\tan\beta<1$ which is generally
not regarded as viable.     
In most of the new physics models considered in the literature the 
enhancement effect
of the single top quark production is smaller than 15\% and some of the models
can even suppress the production rate[8]. An exception to this general result
can exist, as analysed in Ref.[8], in a non-commutting ETC model with
an extra weak gauge boson of mass no less than 500 GeV. 

In this paper we focus on the $R$-parity violating MSSM[11]
and evaluate the effect of $R$-parity violating couplings on single top quark 
production at the Tevatron. While this is an interesting problem in its
own right, the recent anomalous events at HERA provide an additional
motivation for the study of the R-parity violating supersymmetric couplings. 
The HERA data showed excess events in deep-inelastic positron-proton
scattering at high-$Q^2$ and high $x$, which are in apparent conflict with the
Standard Model expectations[12]. The excess events have been interpreted 
as evidence of R-parity breaking supersymmetry[13]. Hence the examination of
effects of R-parity breaking supersymmetry in other processes are desirable.
In Sec.2 we present the Lagrangian for  $R$-parity violating couplings.
In Sec.3 we evaluate the effects of  $R$-parity violating
couplings on the production rate of single top quark.
In Sec.4 we present the numerical results and some discussions.
\vspace{.5cm}

\begin{center} {\Large 2. Lagrangian of  $R$-parity violating couplings}
\end{center}
\vspace{.5cm}

In supersymmetric models, the $R$-parity of a field with spin $S$, 
baryon-number $B$ and  lepton-number $L$ is defined to be
\begin{equation}
R=(-1)^{2S+3B+L}.
\end{equation}
$R$ is +1 for all the SM particles and -1 for all super particles.
$R$-parity invariance is often imposed on the Lagrangian in order to
maintain the separate conservation of baryon-number and lepton-number.
Imposition of $R$-parity conservation has some important consequences;
 super particles must be produced in pairs in collider experiments
and the lightest super particle (LSP) 
must be absolutely stable. Thus the LSP provides a good candidate for cold 
dark matter.

Despite the above mentioned attractive feature of R-parity conservation,
the conservation is not dictated by any fundamental principle such as
gauge invariance and there is no compelling theoretical motivation for it. 
The most general superpotential of 
the MSSM,
consistent with $SU(3)\times SU(2)\times U(1)$
 gauge symmetry and supersymmetry, can be written as
\begin{equation}
{\cal W}={\cal W}_R+{\cal W}_{\not \! R},
\end{equation}
where ${\cal W}_R$ is the $R$-parity conserving part while 
${\cal W}_{\not \! R}$ violates the $R$-parity. They are given by 
\begin{eqnarray}
{\cal W}_R&=&h_{ij}L_iH_2E_j^c+h_{ij}^{\prime}Q_iH_2D_j^c
             +h_{ij}^{\prime\prime}Q_iH_1U_j^c,\\ \label{RV}
{\cal W}_{\not \! R}&=&\lambda_{ijk}L_iL_jE_k^c
+\lambda_{ijk}^{\prime}L_iQ_jD_k^c
             +\lambda_{ijk}^{\prime\prime}U_i^cD_j^cD_k^c+\mu_iL_iH_2.
\end{eqnarray}
Here $L_i(Q_i)$ and $E_i(U_i,D_i)$ are the left-handed
lepton (quark) doublet and lepton (quark) singlet chiral superfields.
$i,j,k$ are generation indices and $c$ denotes a charge conjugate field.
$H_{1,2}$ are the  chiral superfields
representing the two Higgs doublets. 
Note that the term $\mu_i L_iH_2$ can be
rotated away by a redefinition{\footnote{Such redefinition does not leave
the full Lagrangian invariant[14], but it has no relevant 
consequences to our analysis.}} of the Higgs $H_1$ and that of the leptonic
$L_i$ superfields[15] . 

In the  $R$-parity violating superpotential Eq.(\ref{RV}), the  
$\lambda$ and $\lambda^{\prime}$ couplings 
violate lepton-number conservation, while the
$\lambda^{\prime\prime}$ couplings violate baryon-number conservation.
$\lambda_{ijk}$ is antisymmetric in the first two
indices and $\lambda^{\prime\prime}_{ijk}$ is antisymmetric in
the last two indices.
While it is theoretically possible to have both baryon-number 
and lepton-number violating terms in the Lagrangian, the non-observation
of proton decay imposes very stringent conditions on their simultaneous
presence[16].
We, therefore, assume the existence of either  $L$-violating couplings or  
$B$-violating couplings, but not the coexistence of both. 
We calculate the effects of both types of couplings.

In terms of the four-component Dirac notation, the  Lagrangian of the
$\lambda^{\prime}$ and $\lambda^{\prime\prime}$ couplings that 
affect single top production at the Tevatron are given by
\begin{eqnarray}
{\cal L}_{\lambda^{\prime}}&=&-\lambda^{\prime}_{ijk}
\left [\tilde \nu^i_L\bar d^k_R d^j_L+\tilde d^j_L\bar d^k_R\nu^i_L
       +(\tilde d^k_R)^*(\bar \nu^i_L)^c d^j_L\right.\nonumber\\
& &\hspace{1cm} \left. -\tilde e^i_L\bar d^k_R u^j_L
       -\tilde u^j_L\bar d^k_R e^i_L
       -(\tilde d^k_R)^*(\bar e^i_L)^c u^j_L\right ]+h.c.,\\
{\cal L}_{\lambda^{\prime\prime}}&=&-\lambda^{\prime\prime}_{ijk}
\left [\tilde d^k_R(\bar u^i_L)^c d^j_L+\tilde d^j_R(\bar d^k_L)^c u^i_L
       +\tilde u^i_R(\bar d^j_L)^c d^k_L\right ]+h.c.
\end{eqnarray}
The terms proportional to $\lambda$ are not relevant to our present 
discussion and will not be considered here.
As discussed in the proceeding paragraph, we assume that
${\cal L}_{\lambda^{\prime}}$ and ${\cal L}_{\lambda^{\prime\prime}}$  
are not present simultaneously. The Lagrangians given above can give rise to
interesting effects in low energy processes.

In the presence of $R$-parity violation, the phenomenology of 
MSSM changes considerably.
For instance, the LSP is no longer stable and their
decays can lead to interesting phenomenology in collider
experiments. 
Several authors have investigated the phenomenological implications 
of  $R$-parity violating couplings at various colliders[17]. 
Constraints on the R-parity violating couplings have been obtained 
from perturbative unitarity[18,19],
$n-\bar n$ oscillation[19,20], 
$\nu_e$-Majorana mass[21], neutrino-less double $\beta$ decay[22], 
charged current universality[23], $e-\mu-\tau$ universality[23],
$\nu_{\mu}-e$ scattering[23], atomic parity violation[23], 
$\nu_{\mu}$ deep-inelastic scattering[23], $K$-decay[24,25],
$\tau$-decay[26], $D$-decay[26], $B$-decay[27-29] and $Z$-decay at LEP
I[30,31]. As was pointed out in Ref.[24], transforming the Lagrangian from
the gauge basis to the quark mass basis can lead to a flavor changing neutral
current (FCNC) in the up or the down quark sector even under the
assumption of one R-parity violating coupling. FCNC
processes can therefore provide stringent constraints on the R-parity
violating couplings.
\vspace{1cm}

\begin{center} {\Large 3.~ Production rate of single top quark 
                           in  $R$-parity violating MSSM}
		\end{center}
\vspace{.5cm}
The Feynman diagrams for the single top quark production at 
the tree level parton process, $u^i(p_1)+\bar d^j(p_2)\rightarrow t(p_3)
+\bar b(p_4)$, are shown in Fig.1.
The diagrams for the  $L$-violating couplings and  
 $B$-violating couplings are shown in Fig.1(b) and Fig.1(c), respectively.
We assume that figures 1(b) and 1(c) do not exist simultaneously.
Note that the $B$-violating couplings can also give rise to 
a $s$-channel diagram $u^i+d^j\rightarrow \tilde d^k_R \rightarrow t+b$. 
But this is a different process which does not interfere with the SM diagram 
$u^i+\bar d^j\rightarrow W^+ \rightarrow t+\bar b$ and it is relatively
suppressed by the sea structure functions. Therefore, its contribution
is expected to be much smaller at the Tevatron
than the $t$-channel diagram in Fig.1(c).

The amplitudes for Fig.1(a,b,c) denoted by $M^0_{ij}$, 
$\delta M^{\lambda^{\prime}}_{ij}$
and $\delta M^{\lambda^{\prime\prime}}_{ij}$, respectively, are given by 
\begin{eqnarray}
M^0_{ij}&=&i \frac{g^2}{2} \frac{K_{ij}}{\hat s-M_W^2}
  \bar v(p_2) \gamma_{\mu} P_L u(p_1)~  \bar u(p_3) \gamma^{\mu} P_L v(p_4),\\
\delta M^{\lambda^{\prime}}_{ij}&=&-i\lambda^{\prime}_{kij}\lambda^{\prime}_{k33}
\frac{1}{\hat s-M_{\tilde e^k_L}^2+iM_{\tilde e^k_L}\Gamma_{\tilde e^k_L}}
\bar v(p_2)P_Lu(p_1)~\bar u(p_3)P_Rv(p_4),\\
\delta M^{\lambda^{\prime\prime}}_{ij}&=&-i\lambda^{\prime\prime}_{i3k}
\lambda^{\prime\prime}_{3jk}\frac{1}{\hat t-M_{\tilde d^k_R}^2}
\left [\bar u^c(p_1)P_Lv(p_4)-\bar v^c(p_4)P_Lu(p_1)\right ]\nonumber\\
& & \hspace{3cm}\times 
\left [\bar v(p_2)P_Ru^c(p_3)-\bar u(p_3)P_Rv^c(p_2)\right ].
\end{eqnarray}
Here $P_{L,R}\equiv(1\mp \gamma_5)/2$, $\hat t=(p_1-p_4)^2$, 
and the sum over $k=1,2,3$ is implied.
$K_{ij}$ are the KM matrix elements,
 $\hat s$ is the center-of-mass energy squared for the parton-level process,
and $M_{\tilde d^k_R}$ and $M_{\tilde e^k_L}$ are the masses of
squark and slepton, respectively.

The SM parton-level cross section at tree level is 
\begin{equation}
\hat{\sigma}^0_{ij}=\frac{g^4\vert K_{ij}\vert^2}{384\pi}
\frac{(\hat{s}-M^2_t)^2}{\hat{s}^2(\hat{s}-M^2_W)^2}(2\hat s+M^2_t),
\end{equation}
where we have neglected the masses of the bottom quark and the initial partons
{\footnote{We neglected the contribution of third-family sea quark in the 
initial states}}. 
The contribution of  $L$-violating couplings to the parton-level cross section 
is given by
\begin{equation}
\Delta \hat \sigma^{\lambda^{\prime}}_{ij}=\frac{1}{64\pi}
\frac{(\hat{s}-M^2_t)^2}{\hat s}
\left \vert \sum_k \frac{\lambda^{\prime}_{kij}\lambda^{\prime}_{k33}}
{\hat s-M_{\tilde e^k_L}^2+iM_{\tilde e^k_L}\Gamma_{\tilde e^k_L}}
\right \vert^2,
\end{equation}
where $\Gamma_{\tilde e^k_L}$ is the 
width of the charged slepton $\tilde e^k_L$. 
In the R-parity conserving MSSM, the charged sleptons $\tilde e^k_L$ will 
decay into charginos and neutralinos via the processes
$\tilde e^k_L\rightarrow \nu_{e^k}+\bar {\tilde \chi}^+_j$ ($j=1,2$) and
$\tilde e^k_L\rightarrow e^k+\tilde \chi^{0}_j$ ($j=1,2,3,4$), 
where $\tilde \chi^+_j$ and $\tilde \chi^0_j$
represent a chargino and neutralino, respectively[32].
However, in the R-parity violating MSSM, the slepton can also have the
decay modes
$\tilde e^k_L\rightarrow \bar u^j+d^i$ ($i,j=1,2,3$) via the 
$\lambda^{\prime}$ couplings. 
The partial widths are given by
\begin{eqnarray}
\Gamma(\tilde e^k_L\rightarrow \nu_{e^k}+\bar{\tilde \chi}^+_j)
&=&\frac{g^2}{16\pi M_{\tilde e^k}^3}\left \vert U_{j1}\right \vert^2
\left (M_{\tilde e^k}^2-M_{\tilde \chi^+_j}^2\right )^2,\\
\Gamma(\tilde e^k_L\rightarrow e^k+\tilde \chi^{0}_j)
&=&\frac{g^2}{8\pi M_{\tilde e^k}^3}
\left\vert s_W N^{\prime}_{j1}
+\frac{1}{c_W}(\frac{1}{2}-s_W^2)N^{\prime}_{j2}\right\vert^2
\left (M_{\tilde e^k}^2-M_{\tilde \chi^0_j}^2\right )^2,\\
\Gamma(\tilde e^k_L\rightarrow \bar u^j+d^i)
&=&\frac{(\lambda^{\prime}_{kji})^2}{16\pi M_{\tilde e^k}^3}
\left (M_{\tilde e^k}^2-M_{u^j}^2\right )^2,
\end{eqnarray}
where $s_W\equiv \sin\theta_W, c_W\equiv \cos\theta_W$ 
and the masses of the lepton $e^k$ and down-type quark $d^i$ are neglected.
The masses of charginos and neutralinos, and the matrix elements 
$U_{ij}$ and $N^{\prime}_{ij}$ depend on the SUSY parameters $M_2$, $M_1$, 
$\mu$, and $\tan\beta$[10]. 
Here, $M_2$ and $M_1$ are the masses of gauginos corresponding to 
$SU(2)$ and $U(1)$, respectively. $\mu$ is the coefficient of
the $H_1H_2$ mixing term in the superpotential,
and $\tan\beta=v_2/v_1$ is the ratio of the 
vacuum expectation values of the two Higgs doublets.    

The contribution of the $B$-violating couplings to the cross section 
is given by 
\begin{eqnarray}
\Delta \hat \sigma^{\lambda^{\prime\prime}}_{ij}&=&\frac{g^2}{24\pi}
\lambda^{\prime\prime}_{i3k}\lambda^{\prime\prime}_{3jk}K_{ij}
\frac{\hat{s}-M^2_t}{\hat s^2 (\hat s-M_W^2)}
\left [ \hat s+M_t^2-2M_{\tilde d^k_R}^2\right.\nonumber \\
 & & \left. \hspace{2cm} +2M_{\tilde d^k_R}^2
\frac{M_{\tilde d^k_R}^2-M_t^2}{\hat s-M_t^2}
\log \frac{\hat s+M_{\tilde d^k_R}^2-M_t^2}{M_{\tilde d^k_R}^2}\right ],
\end{eqnarray}
where the sum over $k=1,2,3$ is implied.

For the  $B$-violating couplings, we keep only the 
interference term $\delta M_{\lambda^{\prime\prime}} M_0^{\dagger}$.
But for the  $L$-violating couplings, we kept the higher order
term $\vert \delta M_{\lambda^{\prime}}\vert^2$ since 
the interference term $\delta M_{\lambda^{\prime}} M_0^{\dagger}\sim
m_{d^j}m_b$ which drops out when the mass of $d^j$ is neglected.

The total hadronic cross section for the production of single top quark 
is obtained by 
\begin{equation}
\sigma(s)=\sum_{i,j}\int^1_{\tau_0}\frac{d\tau}{\tau}(\frac{1}{s}
\frac{dL_{ij}}{d\tau})(\hat s \hat \sigma_{ij})
\end{equation}
 where $s$ is center-of-mass energy squared,
$\tau_0=(M_t+M_b)^2/s$ and $\tau$ is defined by 
$\tau =x_1x_2$ with $x_{1,2}$ denoting the longitudinal momentum 
fractions of the initial partons $i$ and $j$, respectively. 
The quantity $dL_{ij}/d\tau$ is the parton luminosity defined by
\begin{equation}
\frac{dL_{ij}}{d\tau}=\int^1_{\tau} \frac{dx_1}{x_1}[f^A_i(x_1,\mu)
f^B_j(\tau/x_1,\mu)+(A\leftrightarrow B)]
\end{equation}
where $A$ and $B$ denote the incident hadrons and
the functions $f^A_i$ and $f^B_j$ are the usual parton distributions.
\vspace{1cm}

\begin{center} {\Large 5. Numerical results and conclusion} \end{center}
\vspace{.5cm}

In our numerical calculation, we use the CTEQ3L parton 
distribution functions[33]  with $\mu=\sqrt {\hat s}$, and
assuming $M_t=175$ GeV and  $\sqrt s=2$ TeV. The KM matrix elements 
$K_{12}=-K_{21}=0.22$ are used in our calculation.
\vspace{.5cm}

{\large 5.1 $L$-violating couplings}

For the contribution of  $L$-violating couplings,
we assume the masses of sleptons $\tilde e^k_L$ to be degenerate.
The contribution of  $L$-violating couplings to the cross section is sensitive
to the width of the charged slepton $\tilde e^k_L$, which depends on the SUSY 
parameters $M_2$, $M_1$, $\mu$, $\tan\beta$, and all $\lambda^{\prime}$ 
couplings. 

In our calculation we use the GUT relation 
$M_1=\frac{5}{3}\frac{g'^2}{g^2} M_2\approx \frac{1}{2}M_2$.
Since the masses of charginos and neutralinos are not sensitive to 
$\tan\beta$, we fix $\tan\beta=2$ and retain
$M_2$ and $\mu$ as variables. 

The upper limits of the $L$-violating couplings for the squark mass of
100 GeV are given by
\begin{eqnarray}
\vert \lambda^{\prime}_{kij}\vert&<&0.012,~(k,j=1,2,3; i=1,2),\\
\vert \lambda^{\prime}_{13j}\vert&<&0.16,~(j=1,2),\\
\vert \lambda^{\prime}_{133}\vert&<&0.001,\\
\vert \lambda^{\prime}_{23j}\vert&<&0.16,~(j=1,2,3),\\
\vert \lambda^{\prime}_{33j}\vert&<&0.26,~(j=1,2,3),
\end{eqnarray}
The first set of constraints come from the decay $K\rightarrow \pi \nu
\nu$ with FCNC processes in the down quark sector[24]. The second and forth
set of constraints are obtained from the semileptonic decays of $B$-meson[29]. 
The third constraint,
i.e., that on the coupling $\lambda^{\prime}_{133}$,  is obtained from the 
Majorana mass that the coupling can generate for the electron type neutrino[21].
The last set of limits  have been derived from the leptonic decay modes of 
the $Z$[30].

 Also the following constraints are derived for the combinations
of $\lambda^{\prime}$ couplings
\begin{eqnarray}
&\lambda^{\prime}_{13i}\lambda^{\prime}_{12i},&
\lambda^{\prime}_{23j}\lambda^{\prime}_{22j}<1.1\times 10^{-3}, 
~(i=1,2; j=1,2,3),\\
&\lambda^{\prime}_{i1n}\lambda^{\prime}_{j2n},&
\lambda^{\prime}_{in2}\lambda^{\prime}_{jn1}<10^{-5},~(i,j,n=1,2,3),\\
&\lambda^{\prime}_{111}\lambda^{\prime}_{212},&
\lambda^{\prime}_{112}\lambda^{\prime}_{211},
~\lambda^{\prime}_{121}\lambda^{\prime}_{222}<10^{-7},\\
&\lambda^{\prime}_{122}\lambda^{\prime}_{221},&
\lambda^{\prime}_{131}\lambda^{\prime}_{232},
~\lambda^{\prime}_{132}\lambda^{\prime}_{231}<10^{-7},
\end{eqnarray}
where the first set of constraints are derived in Ref.[29] and the other
derived in Ref.[25].

Inputing the upper limits of the relevant $L$-violating couplings
for squark mass of 100 GeV, 
we get the maximum contribution of  $L$-violating couplings 
which are shown in Fig.2 and Fig.3.
Figure 2 shows the histogram of the differential cross section 
versus the invariant mass of the $t\bar b$ system over a bin size of 
10 GeV with the favorable parameters $M_2=-\mu=200$ GeV.
 The solid line is for the standard model. To illustrate the contributions of
sleptons of different masses, we superpose the effects of three sleptons on
the same curve. The dashed, dotted and
       dash-dotted lines are the standard model plus the slepton
       contributions for three different slepton masses respectively:
       230 GeV, 300 GeV and 350 GeV.
The resonance behavior is already manifested.
Because of their narrow widths, for each slepton the contributions of
the $\lambda^{\prime}$-couplings are negligible for a couple of bins away
from the resonance.  This will help to
identify the signal of the slepton production. 
 
Figure 3 shows the ratio of the total slepton contribution integrated over 
$\hat s$ to that of the standard
model as a function of the slepton mass.
For slepton mass lighter than 180 GeV, the contribution
is very small. When the slepton mass becomes heavier than 180 GeV, which is 
the threshold for the slepton to decay into $t$ and $\bar b$, the contribution
increases sharply and reaches its maximum sizes at the slepton mass of about
230 GeV. Then the effect drops quickly with further increase in
the slepton mass. The large slepton mass suppression can be
understood as follows: When the slepton mass is large the
parton cross section contributions coming mainly from 
$\hat s\sim$mass of slepton require large momenta from the initial partons
which is suppressed by their structure functions. An additional suppression
is caused by the increase of the slepton width when the slepton mass 
increase.

Figure 3 also shows that the contribution to the production rate 
increases with the increase of $M_2$ and $\mu$. 
This is because with the increase of $M_2$ and $\mu$, 
the chargino and neutralino masses increase and thus the width of the slepton  
decreases. The overall contribution to the production rate
can exceed 20\% when slepton mass lies in a narrow range 
( 200 GeV$\sim$ 270 GeV) and $M_2=-\mu>200$ GeV.
As shown in Fig.2, the contribution is confined within a bin of $10\sim 20$GeV. 
So, one can use the upgraded Tevatron to search for enhanced single
top quark production so as
to further constrain the $L$-violating couplings in this mass range 
of sleptons.

\vspace{.5cm}

{\large 5.2 $B$-violating couplings}

In our calculation, we assume the masses of $\tilde d^k_R$ to be degenerate.
Then the contribution of  $B$-violating couplings can be parametrized as
\begin{equation}
\frac{\Delta \sigma_{\lambda^{\prime\prime}}}{\sigma_0}=
 F^{\prime\prime}_{11}\lambda^{\prime\prime}_{132}\lambda^{\prime\prime}_{312}
+F^{\prime\prime}_{12}\lambda^{\prime\prime}_{113}\lambda^{\prime\prime}_{312}
+F^{\prime\prime}_{21}\lambda^{\prime\prime}_{223}\lambda^{\prime\prime}_{312}
+F^{\prime\prime}_{22}\lambda^{\prime\prime}_{213}\lambda^{\prime\prime}_{312},
\end{equation}
where $F^{\prime\prime}_{ij}$ depend on squark mass and are shown in 
Fig.4. From Fig.4 we see that $F^{\prime\prime}_{21}$ and 
$F^{\prime\prime}_{22}$ are the same and negligibly small.
Since $F^{\prime\prime}_{12}$ is smaller than $F^{\prime\prime}_{11}$
and $\lambda^{\prime\prime}_{113}\le 10^{-4}$[19,20], we can also neglect 
the term 
$F^{\prime\prime}_{12}\lambda^{\prime\prime}_{113}\lambda^{\prime\prime}_{312}$.

For squark mass of 100 GeV, we take 
the upper limits for other $B$-violating couplings as given by[18,19,31]
and are derived from perturbative unitarity and $Z$ decay. 
\begin{eqnarray}\label{e20}
\lambda^{\prime\prime}_{312}<0.97,\\  \label{e21}
 \lambda^{\prime\prime}_{132}, \lambda^{\prime\prime}_{223}, 
\lambda^{\prime\prime}_{213}<1.25,
\end{eqnarray}
From these upper limits, we find the upper limit of the contribution
to be $\Delta \sigma^{\lambda^{\prime\prime}}/\sigma^0<2752{\rm \%}$.
This shows that the contribution can be quite large in the allowed 
region of $B$-violating couplings. 
Of course, in this upper limit of couplings, our results are no longer 
reliable and the higher order terms such as 
$(\lambda^{\prime\prime}_{132}\lambda^{\prime\prime}_{312})^2$ 
have to be included. 

Neglecting the terms proportional to $F^{\prime\prime}_{12}$, 
$F^{\prime\prime}_{21}$ and $F^{\prime\prime}_{22}$
we present in Fig.5 the plot corresponding to $\Delta\sigma/\sigma_0=20$ \% 
in the ($\lambda^{\prime\prime}_{132} \lambda^{\prime\prime}_{312}$,
$M_{\tilde q}$) plane.
The region above the plot corresponds to $\Delta\sigma/\sigma_0>20$ \% 
while the region below the plot corresponds to $\Delta\sigma/\sigma_0<20$ \%.
From Fig.5 we see that if we assume an observable level of 20\%
at the upgraded Tevatron, the coupling $\lambda^{\prime\prime}_{132} 
\lambda^{\prime\prime}_{312}$ can be probed down to
0.01, 0.02, 0.03, 0.04, 0.06, 0.08, 0.1 and 0.13 for 
$M_{\tilde q}=$100, 200, 300, 400, 500, 600, 700 and 800 GeV, respectively. 
As mentioned above, the present upper limit is 
$\lambda^{\prime\prime}_{132}\lambda^{\prime\prime}_{312}<1.25\times 0.97
\approx 1.2$ for squark mass of 100 GeV. 
So, the single top production process at the upgraded Tevatron can be 
used meaningfully to probe the product of the $B$-violating 
couplings $\lambda^{\prime\prime}_{132}$ and $\lambda^{\prime\prime}_{312}$. 
\vspace{1cm}
 
In summary, we studied the single top quark production via 
$u \bar d\rightarrow t \bar b$
at the Fermilab Tevatron in $R$-parity violating supersymmetry.
We found that within the allowed range of the coupling constants,
the effects of the $L$-violating $\lambda^{\prime}$ couplings
can be observed at the upgraded Tevatron
only for slepton mass in a narrow range.
As shown in Fig.2 a distinctive resonance is associated with the 
enhanced production of single top. This can also serve as a signal
for the production of slepton via $L$-violating couplings
or further constraint to their coupling strengths. 
For $B$-violating $\lambda^{\prime\prime}$ couplings,
in the allowed range of the relevant coupling constants 
Eqs.(\ref{e20},\ref{e21}), 
the contribution to the production rate can reach the observable level
for a wide range in squark mass. 
So the upgraded Tevatron can make a powerful probe for the product of the
$B$-violating couplings $\lambda^{\prime\prime}_{132}$ and
$\lambda^{\prime\prime}_{312}$. 
Failure to observe a signal of enhancement to the single top production
will be an indication of small $B$-violating couplings as shown
in Fig.5.
We note that owing to the linear vs quadratic dependence
of the products of R-parity violating couplings, the $B$-violating couplings 
will be a more sensitive probe than the $L$-violating couplings. 

It should be noted that the sparticle exchange effect in single top 
quark production explored in the present article is complementary 
to the approach of direct sparticles production at Tevatron.  This
latter approach can also yield a mass limit on the sparticle.  In 
the case of the L-violating 
coupling, a lower limit of 100 GeV on the mass of squark/gluino has 
been obtained in [34] and the limit can possibly be pushed up to 
250 GeV for the accumulated dilepton data from RUN 1 [35].  But this
sparticle production with the L-violating coupling does not allow a 
direct measurement of the slepton mass.  If we use the MSSM with 
grand unification, in which the squark and slepton are related,
a similar limit for the slepton mass can also be obtained as for the 
squark, i.e., a 100/250 GeV mass bound can be set.  The slepton pair
production in R-parity conserving MSSM is also a limited probe of the
slepton mass at the Tevatron [36]. However, our 
approach allows a direct reach of the slepton mass, albeit a limited 
range of values.  For the B-violating SUSY model corresponding to 
the $\lambda''$, the Tevatron reach of squark mass by direct 
production of the squark is very limited.  The approach presented 
in this article can probe a wide range of the
the ratio of the product of the $\lambda''$ and the squark mass.

We conclude by noting that the $s$-channel squark intermediate process,
$u^i+d^j\rightarrow \tilde d^k_R \rightarrow t+b$, is small at the Tevatron
as pointed out early. But this process may not be small in the environment
of $pp$ collision of LHC. This process has an interesting distinctive
signal of double $b$ production and is currently under investigation.
 
\vspace{1cm}

\begin{center}{\Large  Acknowledgement}\end{center}
We would like to thank  X. Tata,  A. P. Heinson,  Z. J. Tao and J. Hauptman 
for helpful discussions and remarks.
This work was supported in part by the U.S. Department of Energy, Division
of High Energy Physics, under Grant No. DE-FG02-94ER40817 and DE-FG02-92ER40730.
XZ was also supported in part by National Natural Science Foundation of China
and JMY acknowledge the partial support provided by the Henan Distinguished
Young Scholars Fund.
\vspace{1cm}

{\LARGE References}
\vspace{0.3in}
\begin{itemize}
\begin{description}
\item [{\rm [1]}]  S. Willenbrock and D. Dicus, Phys. Rev. D34, 155 (1986);\\
                S. Dawson and S. Willenbrock, Nucl. Phys. B284, 449 (1987);\\
                C.-P. Yuan, Phys. Rev. D41, 42 (1990);\\
 	        F. Anselmo, B. van Eijk and G. Bordes, 
                Phys. Rev. D45, 2312 (1992);\\  
                R. K. Ellis and S. Parke, Phys. Rev. D46,3785 (1992);\\
                D. Carlson and C.-P. Yuan, Phys. Lett. B306,386 (1993);\\
                G. Bordes and B. van Eijk, Nucl. Phys. B435, 23 (1995);\\
                A. Heinson, A. Belyaev and E. Boos, hep-ph/9509274. 
\item[{\rm[2]}] S. Cortese and R. Petronzio, Phys. Lett. B306, 386 (1993). 
\item[{\rm[3]}] T. Stelzer and S. Willenbrock, Phys. Lett. B357, 125 (1995). 
\item[{\rm[4]}] A. P. Heinson, hep-ex/9605010. 
\item[{\rm[5]}] M. Smith and S. Willenbrock, hep-ph/9604223;\\
                S. Mrenna and C.-P. Yuan, hep-ph/9703224. 
\item[{\rm[6]}] A. Datta and X. Zhang, Phys. Rev. D55,2530 (1997);\\
                K. Whisnant, J. M. Yang, B.-L. Young and X. Zhang, hep-ph/9702305,
                to appear in Phys. Rev. D. 
\item[{\rm[7]}] C. S. Li, R. J. Oakes and J. M. Yang, Phys. Rev. D55 (1997)1672;\\
	        C. S. Li, R. J. Oakes and J. M. Yang, hep-ph/9611455, 
                to appear in Phys. Rev. D. 
\item[{\rm[8]}] E. H. Simmons, hep-ph/9612402. 
\item[{\rm[9]}] For a review of two Higgs doublet model, 
                see, for example, J. F. Gunion, H. E. Haber, G. L. Kane,
                and S. Dawson, 
                {\bf The Higgs Hunters' Guide}  (Addison-wesley, Teading,
                MA, 1990). 
\item[{\rm[10]}] For reviews of the MSSM, see, for example,
                 H. E. Haber and G. L. Kane, Phys. Rep. 117, 75  (1985);\\
                 J. F. Gunion and H. E. Haber, Nucl. Phys. B272, 1  (1986). 
\item[{\rm[11]}] For reviews, see, for example,
                 D. P. Roy, hep-ph/9303324;\\
                 G. Bhattacharyya, hep-ph/9608415. 
\item[{\rm[12]}] H1 Collab., C. Adloff et al., DESY 97-024;\\
                 Zeus Collab., J. Breitweg et al., DESY 97-025. 
\item[{\rm[13]}]  D. Choudhury and S. Raychaudhuri, hep-ph/9702392;\\
                  G. Altarelli, J. Ellis, G. F. Guidice, S. Lola and M. L. 
                  Mangano, hep-ph/9703276;\\
                  H. Dreiner and P. Morawitz, hep-ph/9703279;\\
                  J. Kalinowski, R. R\"uckl, H. Spiesberger and  P. M. Zerwas,
                  hep-ph/9703288. 
\item[{\rm[14]}] A. Joshipura and M. Nowakowski, Phys. Rev. D51, 5271 (1995);\\
                 F. de Campos,  M. A. Garcia-Jare\~no, A. S. Joshipura,
                 J. Rosiek and J. W. F. Valle, Nucl. Phys. B451, 3 (1995);\\
                 V. Barger, M. S. Berger, R. J. N. Phillips, and T. W\"ohrmann,
                 Phys. Rev. D53, 6407 (1996). 
\item[{\rm[15]}] L. J. Hall and M. Suzuki, Nucl. Phys. B231, 419 (1984). 
\item[{\rm[16]}] C. Carlson, P. Roy and M. Sher, Phys. Lett. B357, 99 (1995);\\
		 A. Y. Smirnov and F. Vissani, Phys. Lett. B380, 317 (1996). 
\item[{\rm[17]}] J. Erler, J. L. Feng and N. Polonsky, Phys. Rev. Lett. 78, 
                 3063 (1997);\\
                 D. K. Ghosh, S. Raychaudhuri and K. Sridhar, hep-ph/9608352;\\
                 D. Choudhury and S. Raychaudhuri, hep-ph/9702392. 
\item[{\rm[18]}] B. Brahmachari and P. Roy, Phys. Rev. D50, 39 (1994). 
\item[{\rm[19]}]J. L. Goity and M. Sher, Phys. Lett. B346, 69 (1995). 
\item[{\rm[20]}] F. Zwirner,  Phys. Lett. B132, 103 (1983). 
\item[{\rm[21]}] S. Dimopoulos and L. J. Hall, Phys. Lett. B207, 210 (1987);\\
                R. M. Godbole, P. Roy and X. Tata, Nucl. Phys. B401, 67 (1993). 
\item[{\rm[22]}] R. N. Mohapatra, Phys. Rev. D34, 3457 (1986);\\
                M. Hirsch, H. V. Klapdor-Kleingrothaus, S. G. Kovalenko,
                Phys. Rev. Lett. 75, 17 (1995);\\
	        K. S. Babu and R. N. Mohapatra, Phys. Rev. Lett. 75, 2276 (1995). 
\item[{\rm[23]}] V. Barger, G. F. Giudice and T. Han, Phys. Rev. D40, 2978 (1989). 
\item[{\rm[24]}] K. Agashe and M. Graesser, Phys. Rev. D54, 4445 (1996). 
\item[{\rm[25]}] D. Choudhury and P. Roy, hep-ph/9603363.
\item[{\rm[26]}] G. Bhattacharyya and D. Choudhury, Mod. Phys. Lett. A10, 
                 1699 (1995). 
\item[{\rm[27]}] D. E. Kaplan, hep-ph/9703347. 
\item[{\rm[28]}] J. Jang, J. K. Kim and J. S. Lee, hep-ph/9701283.
\item[{\rm[29]}] J. Jang, J. K. Kim and J. S. Lee, hep-ph/9704213.
\item[{\rm[30]}] G. Bhattacharyya, J. Ellis and K. Sridhar,
		  Mod. Phys. Lett. A10,1583 (1995). 
\item[{\rm[31]}] G. Bhattacharyya, D. Choudhury and K. Sridhar, Phys. Lett. B355,
                  193  (1995). 
\item[{\rm[32]}] H. Baer, A. Bartl, D. Karatas, W. Majerotto and X. Tata, 
                 Int. J. Mod. Phys. A4, 4111 (1989). 
\item[{\rm[33]}]  H. L. Lai, J. Botts, J. Huston, J. G. Morfin, J. F. Owens,
                  J. W. Qiu, W. K. Tung and H. Weerts, Phys. Rev. D51, 4763 
                  (1995).
\item[{\rm[34]}] D. P. Roy, Phys. Lett. B283, 270 (1992).
\item[{\rm[35]}] H. Baer, C. Kao and X. Tata, Phys. Rev. D51, 2180 (1995);\\
	         M. Guchait and D. P. Roy, Phys. Rev. D54, 3276 (1996).
\item[{\rm[36]}] H. Baer, C.-H. Chen, F. Paige and X. Tata, Phys. Rev. D 49,
                   3283 (1994).
\end{description}
\end{itemize}
\eject

\begin{center}Figure Captions \end{center}

Fig. 1 Feynman diagrams for single top quark production without associated 
jet other than the $b$:  (a) tree-level in the SM,
 (b)  contribution of $L$-violating couplings,  (c)   contribution of
$B$-violating couplings. 
The blobs denote the  $R$-parity violating SUSY vertices. 

Fig. 2 The histogram of the differential cross section 
versus the invariant mass of the $t\bar b$ system over the bin size of 
10 GeV with the parameters $M_2=-\mu=200$ GeV.
The solid line is for the standard model. The dashed, dotted and
       dash-dotted lines are the standard model plus the slepton
       contributions for three different slepton masses respectively:
       230 GeV, 300 GeV and 350 GeV. The vertical scale is in pb/GeV. 

Fig. 3 The integrated contribution of  $L$-violating couplings 
       to the cross section versus the slepton mass.

Fig. 4 The form factors $F^{\prime\prime}_{ij}$ as functions of squark mass. 
      
Fig. 5 The value of $\lambda^{\prime\prime}_{132} \lambda^{\prime\prime}_{312}$
versus squark mass for $\Delta\sigma/\sigma_0=20$ \%. 
The region above the plot corresponds to $\Delta\sigma/\sigma_0>20$ \% 
while the region below to $\Delta\sigma/\sigma_0<20$ \%. 

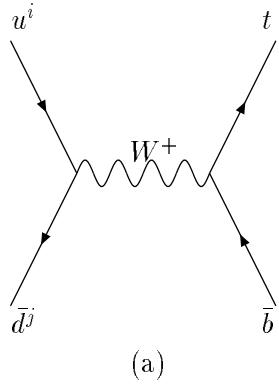
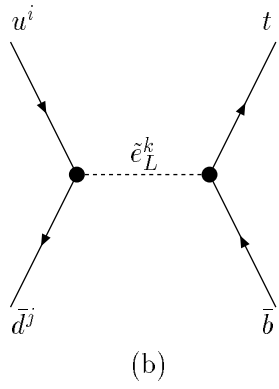
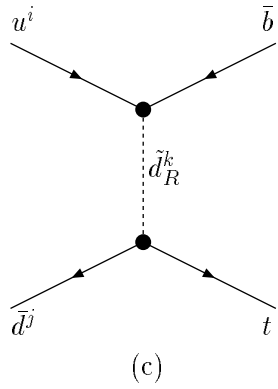
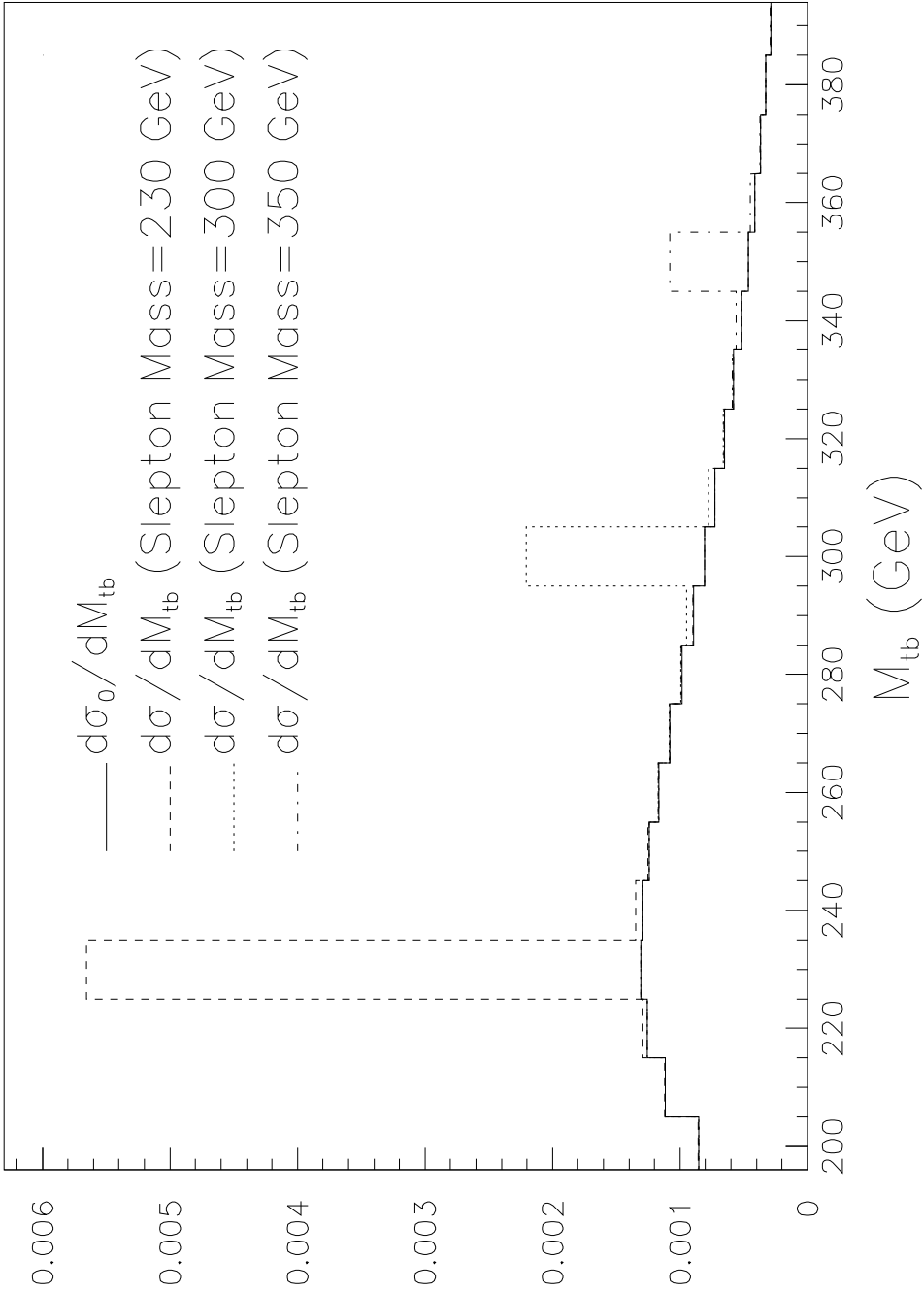
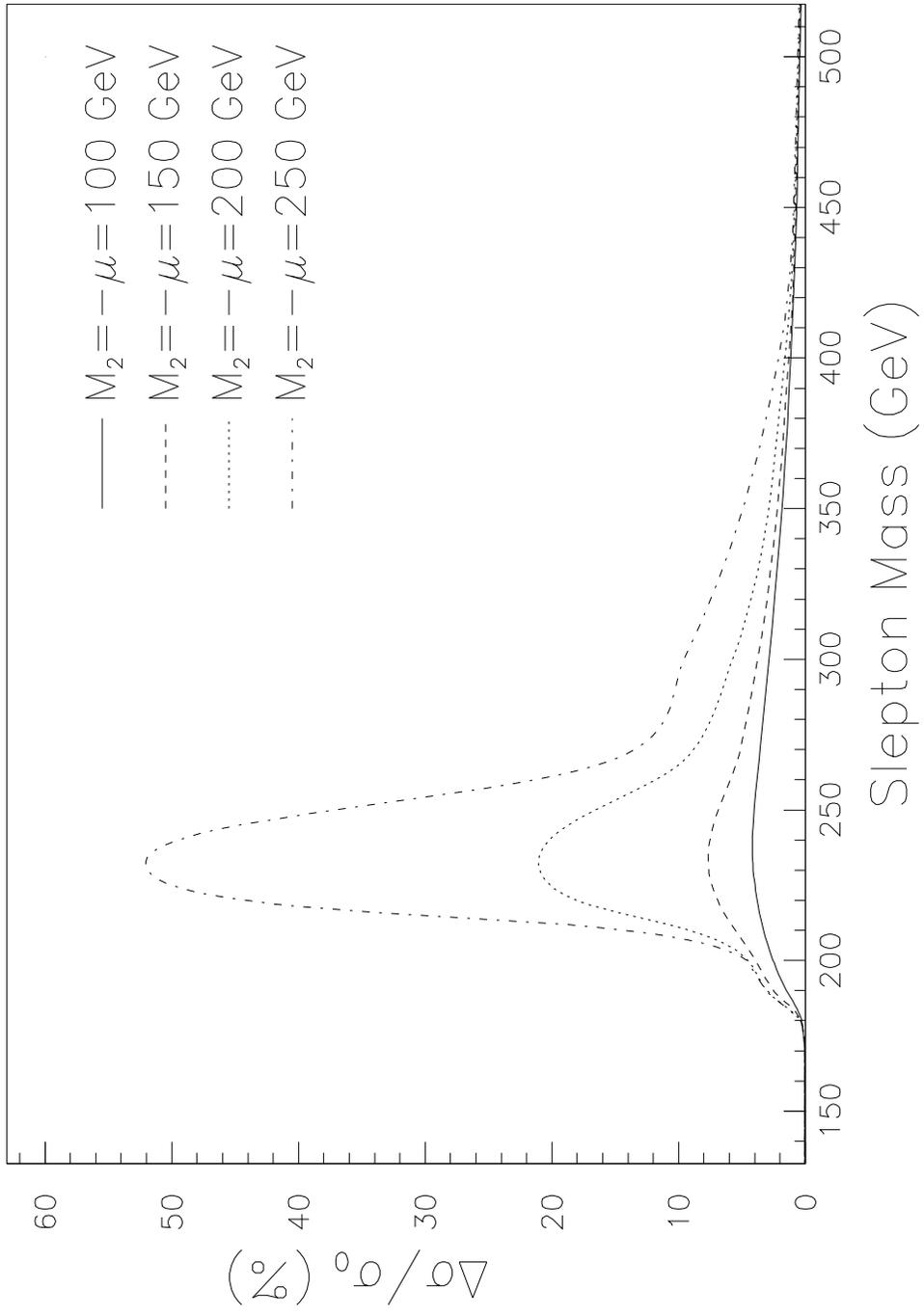
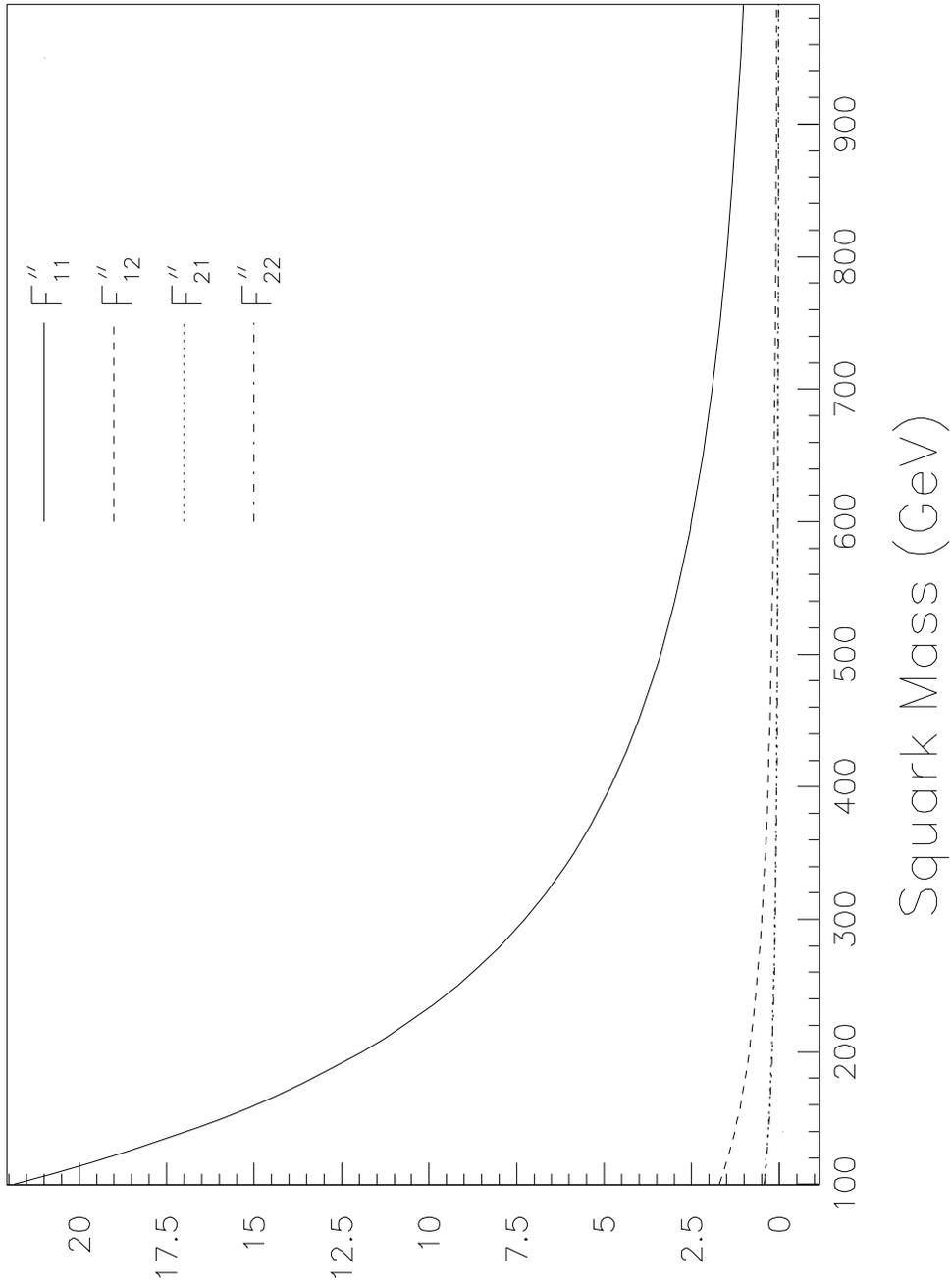
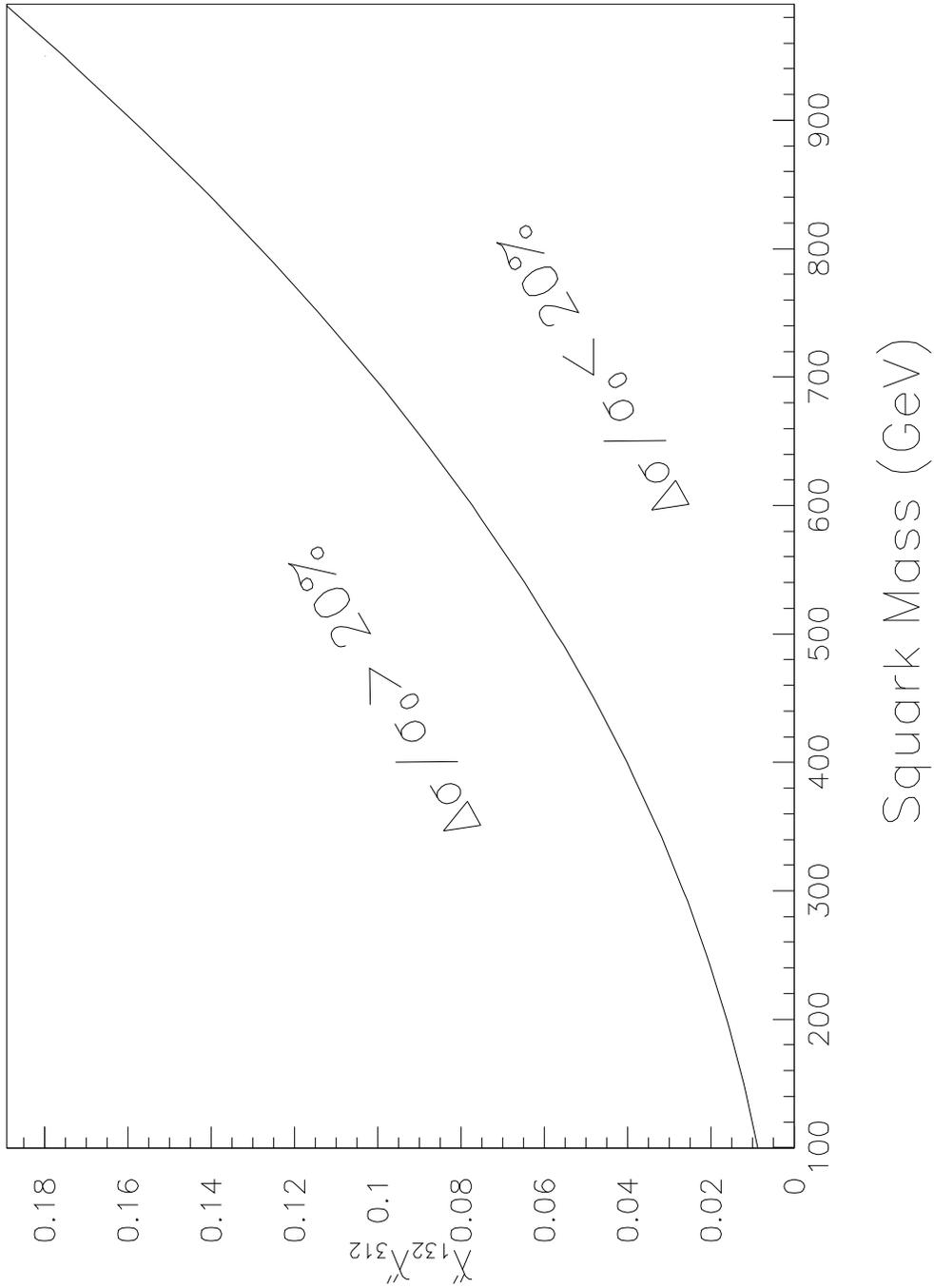
\end{document}